\documentclass[conference]{IEEEtran}
\IEEEoverridecommandlockouts
\usepackage{cite}
\usepackage{amsmath,amssymb,amsfonts}
\usepackage{algorithmic}
\usepackage{graphicx}
\usepackage{textcomp}
\usepackage{xcolor}

\usepackage[utf8]{inputenc} 
\usepackage[T1]{fontenc}    
\usepackage{hyperref}       
\usepackage{url}            
\usepackage{booktabs}       
\usepackage{amsfonts}       
\usepackage{nicefrac}       
\usepackage{microtype}      

\usepackage{xspace}
\usepackage{multirow}   
\usepackage{makecell}   
\usepackage{pifont}
\newcommand{\cmark}{\ding{51}} 
\newcommand{\xmark}{\ding{55}} 
\newcommand{\model}{\texttt{AuroLA}\xspace}
\newcommand{\dataset}{\texttt{AudioVerse}\xspace}

\def\BibTeX{{\rm B\kern-.05em{\sc i\kern-.025em b}\kern-.08em
    T\kern-.1667em\lower.7ex\hbox{E}\kern-.125emX}}
\begin{document}

\title{Re-purposing Multimodal Large Language Models \\ for Audio-Text Retrieval
}

\author{Jilan Xu$^{1}$ ~Carl Thomé$^{2}$ ~Danijela Horak$^{2}$ ~Weidi Xie$^{1,3}$ ~Andrew Zisserman$^{1}$ 
\\
$^1$VGG, University of Oxford
~$^2$Epidemic Sound 
$^3$Shanghai Jiao Tong University
\\
\{jilan,weidi,az\}@robots.ox.ac.uk
~
\{danijela.horak,carl.thome\}@epidemicsound.com
}

\maketitle

\begin{abstract}
Audio–text retrieval is crucial for bridging acoustic signals and natural language. While contrastive dual-encoder architectures like CLAP have shown promise, they are fundamentally limited by the capacity of small-scale encoders. Specifically, the text encoders struggle to understand complex queries that require reasoning or world knowledge. 
In this paper, we propose \textbf{\model}, a novel contrastive language-audio pre-trained model that re-purposes Multimodal Large Language Models (MLLMs) as a unified backbone for audio-text retrieval.
Specifically, we make the following contributions: 
(i) we construct a scalable data pipeline that curates diverse audio from multiple sources and generates multi-granular captions—ranging from long descriptions to structured tags—via automated annotation; 
(ii) we adapt an MLLM for retrieval by prompting it to summarise the audio/text input and using the hidden state of a special token as audio/text embeddings.
(iii) extensive experiments demonstrate that \model consistently outperforms state-of-the-art dual-encoder models, including the recent PE-AV.
This validates the effectiveness of MLLM as a unified backbone for audio-text retrieval.
\end{abstract}


\section{Introduction}
Learning a shared semantic space for acoustic signals and natural language
is a fundamental task in multimodal representation learning since it
 enables bidirectional search between the modalities~\cite{aytar2017see,oncescu2021audio,clap}. 
This capability underpins audio-text retrieval and applications ranging from multimedia indexing~\cite{wold1996content,lallemand2012content,stowell2015detection} to sound-based video search~\cite{vast,valor}. While recent years have seen the dominance of contrastive dual-encoder frameworks like CLAP~\cite{clap}, these approaches are increasingly hitting a scalability ceiling~\cite{af,af2}.

Existing frameworks are primarily limited by two factors: homogeneity of training data, and architectural limits. 
First, training heavily relies on a limited number of sources, for example, AudioSet~\cite{audioset}, restricting audio diversity. The automatically generated captions are either short~\cite{wavcaps} or inclusive of visual information~\cite{soundvecaps,fusionaudio} that cannot be inferred from audio alone.
Second, existing audio encoders and text encoders ({\em e.g.}, HTSAT-RoBERTa~\cite{clap,wavcaps,compa,reclap,flap}) lack the capacity to capture intricate acoustic nuances or to process complex, long-form queries that require compositional reasoning. However, recent breakthroughs in the language and vision community offer a solution: scaling generative Large Language Models (LLMs) has been shown to unlock emergent retrieval capabilities~\cite{e5v,lamra,unime,unime-v2,vlm2vecv2} that significantly outperform traditional dual-tower models~\cite{deepseek,qwen3vlembed,e5v,unime}. These advances suggest that re-purposing generative MLLMs can overcome the limitations of current dual-encoder audio-text retrieval systems.

Driven by this insight, we propose a unified Multimodal Large Language Model (MLLM)-based model for audio–text retrieval named \textbf{\model} (\emph{i.e.}, Aural understanding with Large Language Model). 
\model departs from traditional dual encoder architectures, instead leveraging the reasoning power of MLLMs for retrieval. 
Specifically, we construct a new dataset \textbf{\dataset} by aggregating audio from a diverse range of heterogeneous sources, and develop an automated pipeline capable of generating multi-granular captions that range from high-level semantic tags to detailed, descriptive narratives. This hierarchical supervision is essential for model training, as it equips the MLLM with the ability to recognise both broad acoustic categories and fine-grained temporal details, ensuring robust performance across varied audio search scenarios.

To adapt a generative model for the specific demands of retrieval, we extract audio and text features by prompting the MLLM to generate a compact summary embedding. For effective adaptation, we adopt a two-stage training
strategy to progressively improve the model’s retrieval
performance. The first stage involves training LoRA modules~\cite{lora} via text-only contrastive learning, enabling the MLLM to compress the output sequence into a single embedding vector based on summary prompts. The second stage involves cross-modal LoRA fine-tuning on the constructed multi-granular audio-text dataset, leading to advanced joint audio-text understanding.

Extensive experiments demonstrate that our framework consistently outperforms traditional dual encoders like CLAP and concurrent state-of-the-art models like PE-AV~\cite{peav}, by using only 1.4M audio paired with high quality multi-granular captions.
Dataset annotations, training code and models will be publicly released.

\section{Related Work}

\noindent \textbf{Audio-Text Retrieval~(ATR)} denotes the task of matching an audio clip with its most relevant textual description (or vice versa) from a database of candidates. 
Early works in ATR primarily focused on simpler matching mechanisms, often using structured metadata or short, single-word labels as queries~\cite{foote1997content,slaney2002semantic,helen2007query,chechik2008large}. 
CLAP~\cite{clap} adapted the contrastive learning paradigm from the image domain to the audio domain, setting a precedent for ATR.
Subsequent works extended CLAP by improving audio representations~\cite{flap}, enhancing fine-grained cross-modal alignment~\cite{compa,cacophony}, introducing more robust contrastive objectives~\cite{collat,mltm} or leveraging multilingual textual descriptions~\cite{yan2024bridging, atri}.
Another line of work enhanced the models' audio-text alignment ability by refining existing audio descriptions~\cite{reclap,wavcaps,audiosetcaps,fusionaudio}. 
VAST~\cite{vast}, InternVideo2~\cite{internvideo2} and PE-AV~\cite{peav} further utilised web-scale videos for training joint vision-audio-language models.

\vspace{3pt} \noindent \textbf{Audio-Language Datasets.}
The rapid advancement of audio-text learning tasks largely relies on the availability of datasets with audio class labels or high-quality descriptions. 
For example, LAION-Audio-630K~\cite{clap} and WavCaps~\cite{wavcaps} primarily focused on converting existing tags or web-crawled descriptions, potentially lacking fine-grained acoustic details. 
Later works like Auto-ACD~\cite{autoacd}, SoundVECaps~\cite{soundvecaps} and FusionAudio~\cite{fusionaudio} integrated multimodal information (such as visual objects or places) into the audio descriptions.
Despite the progress, most existing works mainly focused on improving the dataset scale with audio collected from a single data source (\textit{i.e.} AudioSet).
In this work, we aim both to collect audio data from sources beyond AudioSet, and to enhance the quality of textual descriptions in existing datasets, thereby constructing a more diverse and higher-quality audio caption dataset.

\vspace{3pt} 
\noindent \textbf{Multimodal Representation Learning.}
The development of robust and versatile multimodal representations is a foundational challenge in multimodal learning, with the goal of enabling cross-modal understanding~\cite{qwenomni25,worldsense,spotsound,af2} and retrieval~\cite{uniir,gme}. 
Recent studies have incorporated the powerful reasoning and generation capabilities of MLLMs~\cite{uniir,e5v,gme,vlm2vec,lamra,unime}.
E5-V~\cite{e5v} fine-tuned the language component of MLLMs on sentence pairs and demonstrated strong zero-shot retrieval capabilities.
LamRA~\cite{lamra} adopted a retrieval-then-reranking pipeline, progressively
enhancing multimodal retrieval performance. 
VLM2Vec~\cite{vlm2vec,vlm2vecv2} learned multimodal embeddings across diverse
visual forms such as image, video, and visual documents. 
However, existing works primarily focused on learning a shared {\em visual}-text embedding space. 
Inspired by these works, we address the challenging {\em audio}-text retrieval task via MLLMs.

\begin{table*}[t]
\centering
\caption{\textbf{Comparison of source audio--text datasets and \dataset}. Our dataset integrates audio from multiple sources and provides multi-granular textual descriptions.
Average audio length (seconds) and word length are reported with (min$\sim$max).
}
\vspace{-6pt}
\label{tab:dataset_comparison}
\resizebox{0.9\textwidth}{!}{%
\begin{tabular}{lccccccc}
\toprule
\textbf{Dataset} &
\makecell{\#\textbf{Source}} &
\textbf{Scale} &
\textbf{Annotation} &
\makecell{\textbf{Avg. Audio}\\\textbf{Len.}} &
\makecell{\textbf{Avg. Word}\\\textbf{Len.}} &
\makecell{\#\textbf{Caption}/\\\textbf{Audio}} &
\makecell{\textbf{Multi-}\\\textbf{granular}} \\
\midrule
AudioCaps~\cite{audiocaps}
& 1
& 46K
& Human
& 10.0 (1.8$\sim$10)
& 9.0 (2$\sim$39)
& 5
& \xmark \\

Clotho~\cite{clotho}
& 1
& 5K
& Human
& 22.5 (15$\sim$30)
& 11.0 (8$\sim$21)
& 5
& \xmark \\

LAION-Audio-630K~\cite{clap}
& 8
& 630K
& Auto
& 24.6
& 7.3
& 1
& \xmark \\

WavCaps~\cite{wavcaps}
& 4
& 403K
& Auto
& 67.6
& 7.8
& 1
& \xmark \\

Auto-ACD~\cite{autoacd}
& 2
& 1.5M
& Auto
& 10.0
& 18.1
& 1
& \xmark \\

FusionAudio~\cite{fusionaudio}
& 1
& 1.2M
& Auto
& 10.0
& 47.2
& 1
& \xmark \\

SoundVECaps~\cite{soundvecaps}
& 1
& 1.6M
& Auto
& 10.0
& 40.0
& 1
& \xmark \\

AudioSetCaps~\cite{audiosetcaps}
& 1
& 1.9M
& Auto
& N/A
& 28.0
& 1
& \xmark \\

\midrule

\dataset (Ours)
& 11
& 1.4M
& Auto+Human
& 25.9 (0.1$\sim$1800)
& 11.8 (1$\sim$255)
& 3
& \cmark \\

\bottomrule
\end{tabular}%
}
\end{table*}

\begin{figure*}[t]
    \centering
    \includegraphics[width=\textwidth]{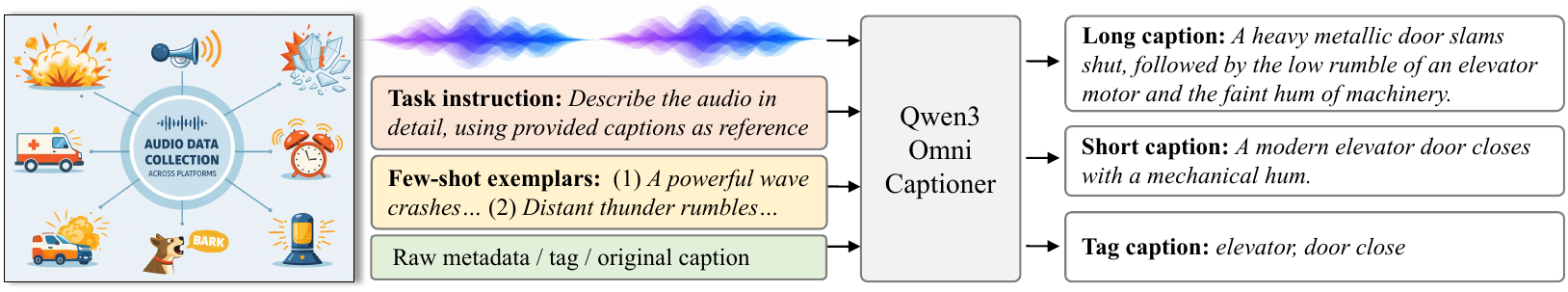}
    \vspace{-16pt}
    \caption{\textbf{Data processing pipeline.} We assemble audio from diverse platforms and datasets. Qwen3-Omni-30B-A3B~\cite{qwenomni3} is used to generate multi-granular captions (including long captions, short captions and tag captions) based on raw audio, task instructions, few-shot examples and auxiliary textual clues. 
    }
    \label{fig:data_pipeline}
    \vspace{-6pt}
\end{figure*}

\section{AudioVerse Dataset}
In this section, we introduce \dataset, an audio–text dataset. 
Our goal is to build a large-scale and diverse audio dataset paired with rich and accurate multi-granular captions, describing acoustic events, sound attributes, and their contextual semantics.

\subsection{Data Source}
\label{subsec:data_source}
To ensure comprehensive coverage of real-world acoustic scenarios, 
we construct the audio-language dataset by aggregating audio data from heterogeneous public resources. This approach contrasts with most existing methods~\cite{audiosetcaps,fusionaudio,soundvecaps}, which primarily rely on AudioSet~\cite{audioset} as their audio source.
Our goal is to capture a diverse spectrum of sound events, encompassing human activities (\emph{e.g.}, conversations, footsteps, cooking sounds), natural soundscapes (\emph{e.g.}, rain, wind, animal vocalizations), and artificial sound effects (\emph{e.g.}, alarms, explosions, machinery). 
We collect and merge existing datasets from 11 sources: AudioSet~\cite{audioset}, FreeSound~\cite{FreeSound}, BBC Sound Effects~\cite{bbc}, Audiostock~\cite{audiostock}, Sonniss Game Effects~\cite{sonniss}, Free to Use Sounds~\cite{ftus}, SoundBible~\cite{soundbible}, VGGSound~\cite{vggsound}, EPIC-Sounds~\cite{epicsound}, AudioCaps~\cite{audiocaps}, and Clotho~\cite{clotho}.
Notably, for academic datasets~\cite{audioset,vggsound,epicsound,audiocaps,clotho} only the training set is selected.
By pooling data from diverse websites and media repositories, we substantially increase both the acoustic diversity and the semantic richness of the dataset, avoiding bias towards a specific data source. 

In addition to raw audio acquisition, we leverage multi-source textual signals associated with each data source. 
This includes original dataset tags or labels~\cite{audioset,vggsound,epidemic,epicsound}, surrounding metadata~\cite{ftus,sonniss}, and language descriptions derived from users or human annotators~\cite{audiocaps,clotho,bbc,audiostock}. The integration of these heterogeneous textual cues enables us to construct high-quality audio-text pairs that go beyond single-label annotations, thereby laying a solid foundation for the multi-granular caption generation pipeline.

\subsection{Multi-granular Caption Annotation}
\label{subsec:data_annotation}
Audio signals naturally encode information at distinct levels of semantic hierarchies. 
A single recording often encompasses fine-grained acoustic nuances ({\em e.g.}, a small dog barks rapidly three times as a car approaches with growing engine noise, and the engine then fades into the distance), mid-level contextual descriptions ({\em e.g.}, a dog barking at a passing vehicle on a quiet street), and high-level categorical labels ({\em e.g.}, dog bark, vehicle engine).
However, prevailing audio-text datasets~\cite{wavcaps,clotho,audiocaps,autoacd,audiosetcaps} generally restrict annotations to a single level of granularity, thereby constraining their use for handling diverse real-world queries. 
Here, we propose a multi-granular captioning framework that assigns complementary semantic descriptions to each audio clip at different levels of abstraction. We generate a triad of annotation types—{\em long captions}, {\em short captions}, and {\em tag captions}—designed to capture comprehensive acoustic details, concise semantic summaries, and structured event concepts, respectively. This multi-layered approach not only enhances audio-text retrieval but also broadens the dataset's applicability to tasks such as audio captioning, tagging, and instruction tuning for MLLMs. As illustrated in Figure~\ref{fig:data_pipeline}, our pipeline is built upon a unified framework using the Qwen3-Omni-30B-A3B audio-language model~\cite{qwenomni3}. 

\vspace{5pt} 
\noindent\textbf{Prompt construction.} 
We employ a composite prompting strategy that feeds the model four inputs: (1) the raw audio signal; (2) a granularity-specific task instruction; (3) few-shot examples demonstrating the desired style; and (4) auxiliary textual cues such as metadata and original labels. This assembly enables robust multimodal reasoning over both acoustic evidence and contextual texts.

\vspace{5pt} 
\noindent\textbf{Multi-granular generation.} 
By fixing the audio and auxiliary cues while varying the instructions, we derive three types of captions from a single framework. 
The model produces: \textbf{long captions}, providing detailed narratives in temporal order; \textbf{short captions}, offering concise summaries of main events; and \textbf{tag captions}, consisting of structured keywords. This design ensures that different semantic granularities are derived from a single, consistent generation framework, avoiding annotation inconsistency while enabling flexible usage across diverse audio understanding tasks.

\subsection{Data Statistics}
\label{subsec:data_statistics}
The dataset comprises 1.4M audio clips collected from 11 heterogeneous sources, see Table~\ref{tab:dataset_comparison}.
The audio duration ranges from very short sound events to long-form recordings, with an average length of 25.9 seconds and a median of 10 seconds, enabling both fine-grained event understanding and broader contextual reasoning.
A key distinction of our dataset is that it is the only large-scale dataset providing multi-granular captions. 
On average, long/short/tag captions contain 21.6/9.7/4.3 words, respectively, forming a hierarchical supervision structure that supports diverse retrieval scenarios.
To structure the tag space, we apply K-means clustering to all tags based on tag features extracted by Qwen3-Embedding~\cite{qwen3}, and obtain 1,200 semantic clusters. Frequently appearing tags include \textit{man speaks}, \textit{engine}, \textit{water}, \textit{vehicle}, \textit{guitar}, reflecting the broad coverage of human activities, natural soundscapes, and artificial sound events.

\subsection{Quality Assessment}
\label{subsec:quality_assessment}
We randomly sample 500 audio clips from \dataset for quality assessment. Three experts, with over five years' experience in machine learning and audio processing, then rated the long, short, and tag captions on a 1–10 scale. 
Long captions were evaluated on event accuracy, completeness, temporal consistency, and acoustic detail, while short and tag captions were assessed on event accuracy and completeness. 
The average scores (± standard deviation) were 8.06 (±2.03), 8.44 (±1.95), and 8.26 (±2.09) for long, short, and tag captions, respectively, demonstrating consistently high semantic quality across all caption types.

\begin{figure*}
    \centering
    \includegraphics[width=1.0\linewidth]{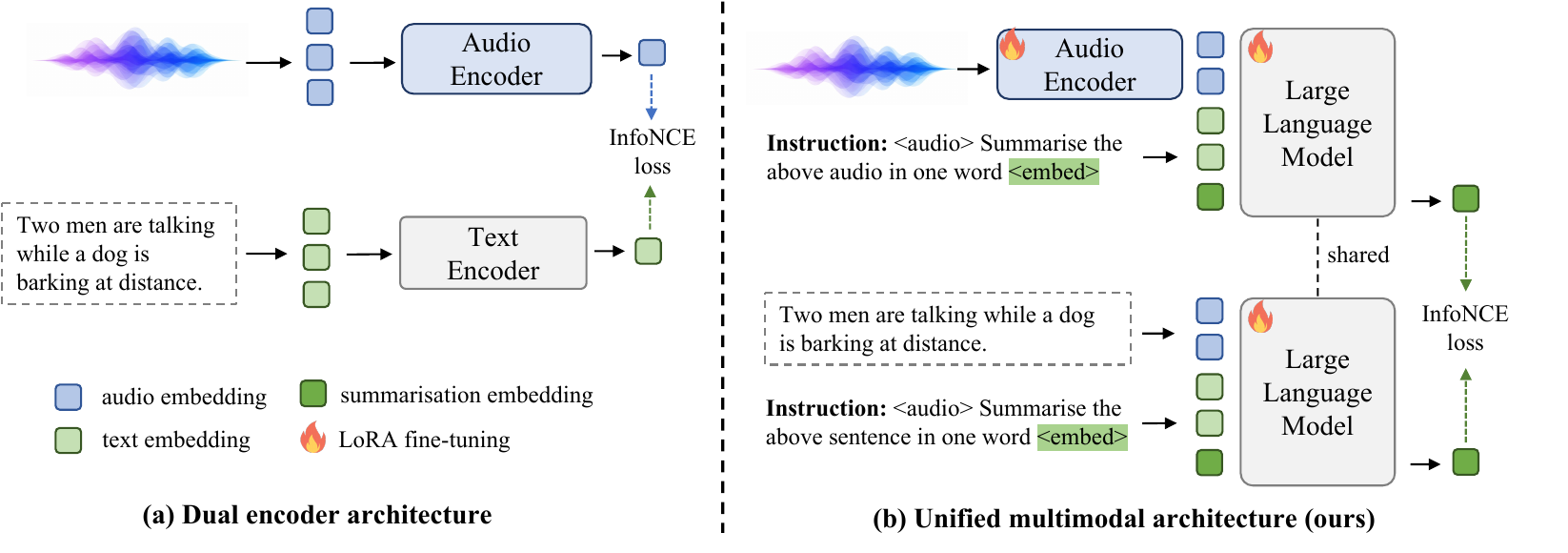}
    \vspace{-20pt}
    \caption{\textbf{Architecture comparison} between the classic dual encoder model (left) and our unified MLLM-based retrieval model (right). The retrieval model is trained by aligning the embedding tokens of audio and text inputs via contrastive learning.}
    \vspace{-6pt}
    \label{fig:architecture}
\end{figure*}

\section{AuroLA: learning from MLLMs}
In this section, we first present a formal definition of the audio-text retrieval task in Sec.~\ref{subsec:problem_formulation}, then we introduce the model architecture in Sec.~\ref{subsec:architecture}.
Lastly, we detail the training and inference pipeline in Sec.~\ref{subsec:pipeline}. 

\subsection{Problem formulation}
\label{subsec:problem_formulation}
In this paper, we study the audio-text retrieval task, which aims to find semantically relevant text samples given a query audio, or vice versa, by measuring similarity in a shared embedding space. We  will use the audio-to-text retrieval task as the example to illustrate the problem formulation.
Formally, let $a$ denote an audio query, and $\Omega=\{t_1,t_2,t_3,...,t_M\}$ denote a candidate text set of size $M$, where each candidate $t_i$ is a natural language description of the sound events in the audio. The objective of audio-text retrieval is to rank all candidates in $\Omega$ according to their semantic relevance to the query audio $a$, and return the top-K most relevant items. 

The dominant approaches~\cite{wavcaps,clap,reclap} encode audio and text pairs into a shared multimodal feature space using separate encoders $\phi_\text{audio}$, $\phi_\text{text}$ (Fig.~\ref{fig:architecture} left), and measure the similarity between audio and text embeddings. In contrast, we leverage one unified multimodal model $\phi_{\text{embed}}$. Specifically, given a pair of query and candidate ($a$,$t$), their similarity is calculated as:
\begin{equation}
    s(a, t) = \text{sim}(\phi_{\text{embed}}(a),\phi_{\text{embed}}(t)),
\end{equation}
where $\text{sim}(\cdot)$ refers to a similarity function, such as cosine similarity or dot product. 
Then, the top-K candidates are selected by ranking all candidates based on their similarity score:
\begin{equation}
    \Omega_K=\Phi_{\text{ret}}(a, \Omega)=\text{Top-K}_{t_i\in\Omega}  s(a, t_i).
\end{equation}
This dense retrieval stage enables efficient large-scale retrieval using approximate nearest neighbor (ANN) search. 
Next, we detail our unified architecture for this retrieval task.

\subsection{Architecture}
\label{subsec:architecture}
The overall architecture of our retrieval model is illustrated in Figure~\ref{fig:architecture} (right). 
Our unified model, denoted as $\phi_{\text{embed}}$, is built upon a Multimodal Large Language Model~\cite{qwenomni25,qwenomni3}, consisting of 
an audio encoder $\phi_{\text{audio}}$ and a large language model $\phi_{\text{llm}}$.
To transform the generative model into an embedding model, we adopt the Explicit One-word Limitation approach~\cite{e5v,lamra}. In the following sections, we detail the encoding process for both the input audio and the corresponding caption.

\vspace{3pt} \noindent\textbf{Audio encoding.} 
Given a raw audio waveform $a_i$, we perform resampling with a frequency of 16,000 Hz, and then transform the resampled waveform into a 128-channel mel-spectrogram with a window size of 25ms and a hop size of 10ms. 
The audio encoding process is defined as:
\begin{equation}
\mathbf{a}_i
= \phi_{\text{llm}}\!\left(
  \left[
    (\phi_{\text{audio}}(a_i),
    \textbf{Prompt}_{\text{A}},
    \texttt{<embed>}
  \right]
\right).
\end{equation}
where $[]$ denotes concatenation. The audio encoder~\cite{qwenaudio} consists of a stack of Transformer encoder layers, followed by a projection layer.
We use the audio prompt $\textbf{Prompt}_\text{A}$ as follows: 
\begin{equation}
\small
    \texttt{<audio> Summarise the above audio in one word:} 
\end{equation}
\texttt{<audio>} is a placeholder to be replaced by audio features as input to the language model. 
Notably, a special \texttt{<embed>} token is appended at the end of the sentence.  
The language model $\phi_{\text{llm}}$ takes the audio features and word embeddings and predicts the output token embeddings autoregressively. 
We simply take the $\texttt{<embed>}$ token embedding as the final audio representation $\mathbf{a}_i$.

\vspace{2pt}
\noindent\textbf{Language encoding.} 
For the text branch, we obtain the text representation by encoding the original caption $t_i$ followed by the text prompt $\textbf{Prompt}_\text{T}$ and the \texttt{<embed>} token:
\begin{equation}
    \textbf{t}_i=\phi_{\text{llm}}([t_i, \textbf{Prompt}_\text{T}, \texttt{<embed>}])
\end{equation}
$\textbf{Prompt}_\text{T}$ is similarly defined as: 
\texttt{Summarise the above text in one word:}. 
Here, the text input $t_i$ is randomly sampled from long captions, short captions, and semantic tags.

\vspace{2pt}
\noindent\textbf{Objective Function.}
To train the generative embedding model based on the obtained audio representation $\textbf{a}_i$ and text representation $\textbf{t}_i$, we employ the contrastive learning paradigm to align paired audio-captions in the embedding space, while pushing unpaired audio and text samples further away. 
The widely adopted objective function for contrastive learning is InfoNCE~\cite{clip,cpc}.
Given a batch of $N$ samples, the audio-to-text contrastive loss is calculated as:
\begin{equation}
   \mathcal{L}_{\text{InfoNCE}} =  - \frac{1}{N} \sum_{i=1}^{N} \log \frac{ e^{s(\mathbf{a}_i, \mathbf{t}_i) / \tau }}{ \sum_{j=1}^{N}e^{ s(\mathbf{a}_i, \mathbf{t}_j) / \tau}
} 
\end{equation}
where $\tau$ is the temperature parameter. The text-to-audio contrastive loss can be calculated similarly, and we take the average as the final objective.

\subsection{Training and Inference Pipeline}
\label{subsec:pipeline}
We adopt Qwen2.5-Omni-7B~\cite{qwenomni25} as the baseline model. 
The training is conducted in two stages: 
(1) First, we adapt the generative multimodal model to the embedding-based retrieval task by text-only contrastive learning on the Natural Language Inference (NLI~\cite{nli}) dataset. 
The model is trained with LoRA~\cite{lora} modules in the LLM.
(2) Next, we further fine-tune the LoRA adapted model by audio-text contrastive learning on the \dataset dataset. 
In this stage, we add LoRA modules to both the audio encoder and LLM. 
At inference time, we extract normalised audio and text embeddings using the retrieval model and calculate the audio-text similarity matrix.

\section{Experiments}

\subsection{Datasets \& Evaluation Metrics}
To comprehensively evaluate audio-text retrieval, 
we adopt AudioCaps~\cite{audiocaps}, Clotho~\cite{clotho}, and Auto-ACD~\cite{autoacd} as the evaluation benchmarks. 
Following prior works~\cite{clap,wavcaps,autoacd,reclap}, 
we adopt Recall@1 as the main evaluation metric. 
AudioCaps and Clotho have five captions per audio. The audio-text retrieval (A2T) is evaluated by identifying the highest-ranked text for each audio query. Conversely, for text-audio (T2A) retrieval, the performance metrics are averaged across all individual text queries to ensure a robust assessment.

\subsection{Implementation Details}
Our model is initialised from Qwen2.5-Omni-7B~\cite{qwenomni25}. 
In Stage-1 text-only contrastive learning, the model is trained for 2 epochs on the Natural Language Inference (NLI) dataset~\cite{nli} with a batch size of 576 and learning rate of 2e-4. 
In Stage-2 audio-text training stage, the model is trained for 2 epochs with a total batch size 512. The initial learning rate is set to 1e-4 with a cosine learning rate decay.
The LoRA rank and alpha are set to 128 and 256, respectively.
The experiments are conducted on 8 NVIDIA H200 GPUs.

\begin{table}[t]
\centering
\caption{\textbf{Main results on AudioCaps~\cite{audiocaps}, Clotho~\cite{clotho}, and Auto-ACD~\cite{autoacd}}. PT stands for pre-training without downstream training sets. * refers to our reproduced results.}
\vspace{-6pt}
\label{tab:main_exp}
\resizebox{\columnwidth}{!}{
\begin{tabular}{l cc cc cc}
\toprule
\multirow{2}{*}{\textbf{Method}} 
& \multicolumn{2}{c}{\textbf{AudioCaps}} 
& \multicolumn{2}{c}{\textbf{Clotho}} 
& \multicolumn{2}{c}{\textbf{Auto-ACD}}  
\\
\cmidrule(lr){2-3}
\cmidrule(lr){4-5}
\cmidrule(lr){6-7}
& T2A & A2T
& T2A & A2T
& T2A & A2T
\\
\midrule
OnePeace (PT)* \cite{onepeace} 
& 20.7 & 24.0 
& 11.1 & 16.2 
& 21.7 & 24.2 
\\
VAST (PT)* \cite{vast} 
&  25.4 & 34.8 
&  16.1 & 20.0 
&  26.4 & 27.5 
\\
PE-AV (PT) \cite{peav} 
&  33.7 & 48.5
&  17.5 & 26.3
&  32.9 & 33.2 
\\
\textbf{\model (PT)} 
& \textbf{42.1}	& \textbf{54.2}
& \textbf{25.1}	& \textbf{32.5}
& \textbf{39.8}	& \textbf{37.5 }
\\
\midrule
CLAP \cite{clap} 
&  32.7 & 43.9 
&  15.6 & 23.7
&  -- & --
\\
CLAP (fusion) \cite{clap} 
&  36.2 & 45.0 
&  17.2 & 24.2 
&  17.9 & 20.0 
\\
DiffATR \cite{diffatr} 
&  36.1 & 42.6 
&  16.7 & 18.8 
&  -- & -- 
\\
CompA-CLAP \cite{compa} 
&  36.1 & 45.2 
&  16.8 & 19.7 
&  -- & -- 
\\
ReCLAP \cite{reclap} 
&  37.1 & 48.0 
&  18.9 & 20.5 
&  -- & -- 
\\
M-LTM \cite{mltm} 
&  39.1 & 49.9
&  16.6 & 22.1 
&  -- & -- 
\\
WavCaps \cite{wavcaps} 
&  39.7 & 51.7 
&  19.5 & 23.4 
&  27.0 & 28.3 
\\
FLAP (fusion) \cite{flap} 
&  41.5 & 53.0 
&  20.3 & 25.5 
&  -- & -- 
\\
Cacophony \cite{cacophony} 
&  41.0 & 55.3 
&  20.2 & 26.5 
&  -- & -- 
\\
ML-CLAP \cite{yan2024bridging} 
&  40.4 & 55.7 
&  23.6 & 29.3 
&  -- & -- 
\\
OnePeace \cite{onepeace} 
&  42.5 & 51.0
&  22.4 & 27.1
&  27.9 & 30.8 
\\
PE-AV \cite{peav} 
&  45.8 & \textbf{63.3}
&  23.0 & 32.7
&  -- & -- 
\\
\midrule
\textbf{\model} 
& \textbf{46.0}	& 63.1	
& \textbf{25.3}	& \textbf{37.0}
& \textbf{39.9}  & \textbf{38.6}
\\
\bottomrule
\end{tabular}
}
\end{table}

\subsection{Comparison with State-Of-The-Art}

The comparison with SoTAs is shown in Table~\ref{tab:main_exp}. 
For fair comparison with prior works~\cite{wavcaps,peav}, 
we report results on: (i) a Pre-Training (PT) setting where training sets of downstream benchmarks (\emph{i.e.}, AudioCaps and Clotho) are excluded from the pre-training, and (ii) a model trained on the full \dataset dataset.
We make the following observations:
1) As reported in the upper half of Table~\ref{tab:main_exp}, our approach achieves state-of-the-art performance across all benchmarks \textbf{under the PT setting}. 
\model (PT) exhibits large gains over PE-AV (PT) despite using only a fraction of their data (1M vs 92M). In particular, on AudioCaps, \model outperforms PE-AV and on T2A (33.7 vs 42.1) and A2T (48.5 vs 54.2).
(2) \textbf{With full \dataset training} illustrated in the lower half of Table~\ref{tab:main_exp}, \model still maintains a consistent advantage over PE-AV, for example, on Clotho, we achieve improvements on T2A (23.0 vs 25.3) and A2T (32.7 vs 37.0).
These results highlight the effectiveness of our proposed unified MLLM backbone in enhancing general retrieval performance over existing dual-encoder models. 
Please refer to the arxiv version for more results on R@5 and R@10.

\begin{table}[t]
\centering
\caption{\textbf{Ablation study} on AudioCaps~\cite{audiocaps} and Clotho~\cite{clotho}.}
\vspace{-6pt}
\label{tab:ablation}
\begin{tabular}{lcccc}
\toprule
\multirow{2}{*}{\textbf{Method}}
& \multicolumn{2}{c}{\textbf{AudioCaps}} 
& \multicolumn{2}{c}{\textbf{Clotho}} \\
\cmidrule(lr){2-3} \cmidrule(lr){4-5}
& T2A & A2T & T2A & A2T  \\
\midrule
Baseline 
& 14.1 & 14.5 & 11.1 & 9.7 \\
+ Stage-1 text-only pre-training
& 22.0 & 27.2 & 19.0 & 18.4  \\
+ Stage-2 audio-text training
& 31.9 & 42.7 & 21.1 & 25.3  \\
+ Multi-granular caption
& 40.4 & 49.4 & 25.5 & 32.5 \\
\bottomrule
\end{tabular}
\end{table}

\begin{table}[t]
\centering
\caption{\textbf{Comparison of different pre-training datasets} under different backbones. For fair comparison, the training sets of AudioCaps and Clotho are excluded from the pre-training data.}
\vspace{-6pt}
\label{tab:pretrain_dataset_compare}
\resizebox{\columnwidth}{!}{%
\begin{tabular}{lccccccc}
\toprule
\multirow{2}{*}{\textbf{Dataset}} & \multirow{2}{*}{\textbf{Scale}} 
& \multicolumn{2}{c}{\textbf{AudioCaps}} 
& \multicolumn{2}{c}{\textbf{Clotho}} 
& \multicolumn{2}{c}{\textbf{Auto-ACD}} \\
\cmidrule(lr){3-4} \cmidrule(lr){5-6} \cmidrule(lr){7-8}
& & T2A & A2T & T2A & A2T & T2A & A2T \\
\midrule
\small{\textit{\textcolor{gray}{CLAP as backbone}}} \\
WavCaps~\cite{wavcaps} 
& 0.4M 
& 28.6 & 40.2 & 16.5 & 20.0 & 24.5 & 25.6\\

AudioSetCaps~\cite{audiosetcaps}
& 1.9M 
& 32.8 & 44.5 & 10.5 & 14.4 & 29.1 & 30.5 \\

\textbf{\dataset} 
& 1.3M 
& \textbf{40.9}	& \textbf{52.6}	& \textbf{16.8} & \textbf{23.3} & \textbf{32.7} & \textbf{34.5}	\\
\midrule

\small{\textit{\textcolor{gray}{AuroLA as backbone}}} \\
WavCaps~\cite{wavcaps}
& 0.4M 
& 31.9 & 42.7 & 21.1 & 25.3 
&26.5	&26.6\\

FusionAudio~\cite{fusionaudio}
& 1.2M & 
31.3 & 45.0		& 	19.8	& 		25.4	& 		32.9		& 	\textbf{39.1}  \\

AudioSetCaps~\cite{audiosetcaps}
& 1.9M 
&38.6	&47.7	&21.0	&27.2	&35.4	&37.2	\\

\textbf{\dataset}
& 1.3M 
& \textbf{42.1} & \textbf{54.2} & \textbf{25.1} & \textbf{32.5}
&\textbf{39.8}	&37.5 \\
\bottomrule
\end{tabular}}%
\end{table}

\subsection{Ablation Study and Discussions}
In this section, we study the main components of our approach. 
Unless otherwise specified, we adopt a subset of \dataset for training, 
containing 373k audio-caption pairs~\cite{wavcaps} sourced from AudioSet, 
BBC Sound Effect, Freesound and SoundBible. 

\vspace{4pt} \noindent\textbf{Main ablation.}
Table~\ref{tab:ablation} presents a systematic ablation study on the AudioCaps~\cite{audiocaps} and Clotho~\cite{clotho} benchmarks to quantify the contribution of each core component in our framework, from the following aspects:
\textbf{(i) baseline performance.} Our baseline model is the vanilla Qwen-Omni-2.5-7B~\cite{qwenomni25} without additional training. We observe consistently low zero-shot retrieval performance across both text-to-audio (T2A) and audio-to-text (A2T) settings. This is expected, as the model is natively optimised in an autoregressive manner for generative tasks ({\em e.g.}, QA, multi-turn dialogue) rather than discriminative embedding-space alignment;
\textbf{(ii) impact of multi-stage training.} 
Introducing Stage-1 text-only pre-training  leads to a substantial performance boost on both datasets. This indicates that large-scale text-text contrastive learning effectively ``warms up'' the MLLM's latent space for retrieval-based tasks. 
Building upon this, Stage-2 training on audio-caption pairs further improves retrieval accuracy—particularly for A2T retrieval, thereby successfully adapting the model to bridge the modal gap between audio representations and textual semantics;
\textbf{(iii) effect of multi-granular descriptions}.
Next, we replace the original dataset captions with our refined multi-granular descriptions. These refined captions yield significant gains on both datasets, suggesting that richer, more structured textual supervision effectively enhances semantic consistency across modalities and provides the model with more discriminative features.

\vspace{4pt} \noindent\textbf{Analysis on \dataset dataset.}
To systematically evaluate the impact of different pre-training corpora, we conduct a controlled comparison by fixing the model backbone and training with the InfoNCE loss~\cite{cpc} across several prominent audio–text datasets, including WavCaps~\cite{wavcaps}, AutoACD~\cite{autoacd}, AudioSetCaps~\cite{audiosetcaps}, and FusionAudio~\cite{fusionaudio}. To ensure a rigorous zero-shot evaluation, we exclude the training sets of AudioCaps and Clotho from all pre-training corpora. 
As shown in Table~\ref{tab:pretrain_dataset_compare}, although \dataset is not the largest in scale compared to recent massive datasets like AudioSetCaps or AutoACD, it consistently delivers superior downstream retrieval performance. This advantage is primarily driven by the acoustic diversity of our data and the richness of our supervision. Specifically, while prior approaches~\cite{autoacd,fusionaudio,audiosetcaps} often rely on limited sources such as AudioSet or VGGSound—which can induce distribution biases toward specific downstream tasks—\dataset aggregates audio from a much broader range of sources to ensure comprehensive coverage of real-world sound events. 
Complementing this diversity, our automated pipeline generates multi-granular captions, including long-form sentences, short descriptions, and semantic tags; this provides more structured supervision than the single caption format used in previous works, thereby mitigating modality misalignment. In addition, the effectiveness of \dataset generalizes across architectures, yielding consistent gains on both dual encoders (\emph{e.g.} CLAP~\cite{clap}) and MLLM-based backbones in T2A and A2T settings. 
These results demonstrate that multi-source diversity and multi-granular supervision are as critical as raw data scale for learning robust, transferable audio–text representations. 

\begin{figure}[t]
    \centering
    \includegraphics[width=0.9\columnwidth]{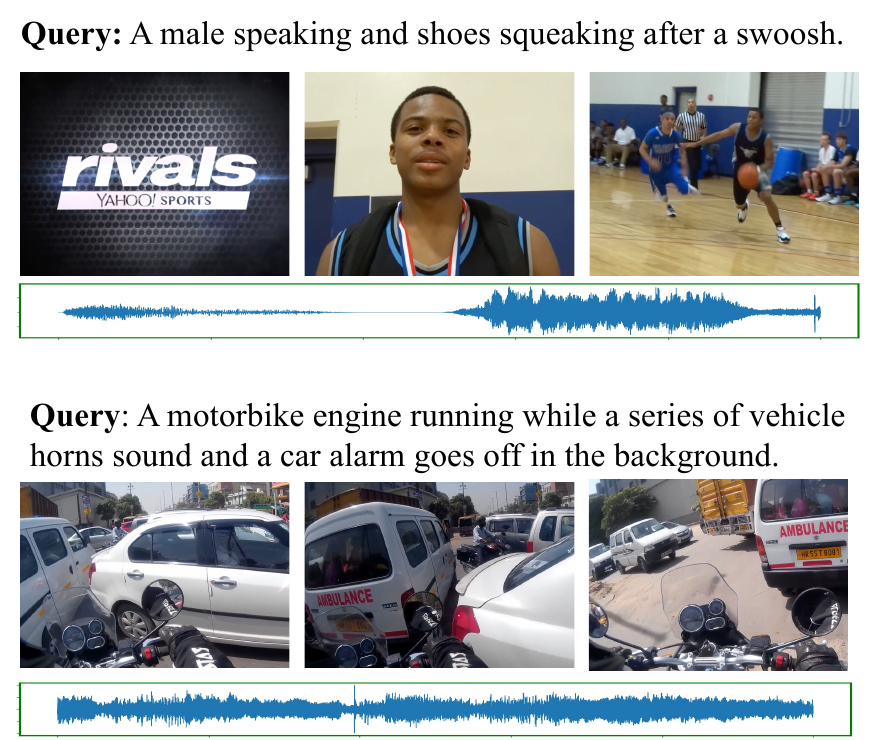}
    \vspace{-6pt}
    \caption{\textbf{Visualisation of text to audio retrieval on AudioCaps}. Visual information is only for reference. 
    }
    \label{fig:visualisation_t2a}
    \vspace{-6pt}
\end{figure}

\begin{table}[t]
  \centering
  \caption{\textbf{Inference cost comparison} on AudioCaps using a H200 GPU. We report Latency (ms\,/\,sample), Peak Mem (GB) and GFLOPs.}
  \vspace{-6pt}
  \label{tab:inference-cost}
  \resizebox{\linewidth}{!}{%
  \begin{tabular}{l c c c c c}
  \toprule
  \multirow{2}{*}{\textbf{Model}}
    & \multirow{2}{*}{\textbf{Latency}}
    & \multirow{2}{*}{\textbf{Peak Mem}}
    & \multirow{2}{*}{\textbf{GFLOPs}} 
    & \multicolumn{2}{c}{\textbf{Recall@1}} \\
  \cmidrule(lr){5-6}
  & & & & T2A & A2T \\
  \midrule
   CLAP~\cite{clap} & 1.2  & 2.2  & 10.5  & 32.7  & 43.9 \\
   AuroLA (ours)    & 37.2 & 18.0 & 7.8k & 46.0 & 63.1 \\
  \bottomrule
  \end{tabular}%
  }
  \vspace{-15pt}
  \end{table}
\vspace{4pt} \noindent\textbf{Inference cost comparison.}
Table~\ref{tab:inference-cost} contrasts AuroLA against a lightweight CLAP baseline along three efficiency axes.
Despite a higher inference cost, AuroLA attains stronger retrieval accuracy (32.7 vs 46.0). 
This disparity reflects a fundamental capacity–efficiency trade-off between contrastive dual-encoders and unified MLLM-based retrieval models.

\vspace{4pt} \noindent\textbf{Qualitative analysis.}
In Figure~\ref{fig:visualisation_t2a}, we visualise the Top-1 retrieved audio clip for given text queries on AudioCaps~\cite{audiocaps}. 
Here, the YouTube video frames corresponding to audio are only for reference.
We observe that \model consistently retrieves acoustically and semantically aligned samples. 
For instance, when the text query involves composite scenes (e.g., “motorbike engine and car alarm”), the retrieved audio exhibit consistent temporal patterns and reverberation characteristics, suggesting that the model captures multi-source acoustic composition rather than relying on a single dominant cue. 
Meanwhile, samples containing a single acoustic cue (e.g. “shoes squeaking sound in the basketball court”) are also retrieved with high similarity. 

\section{Conclusion}
In this work, we presented \model, a scalable and unified model that re-purposed Multimodal Large Language Models as the core backbone for audio–text retrieval. 
Our approach leveraged a curated, heterogeneous dataset \dataset and an automated multi-granular annotation pipeline, which provides the model with a rich semantic hierarchy of long captions, short descriptions, and structured tags.
By training on \dataset, \model consistently surpassed strong concurrent works across multiple audio-text retrieval benchmarks, 
validating the effectiveness of transitioning from specialized dual encoders to unified MLLM-centric architectures for audio–text retrieval.
We believe this work takes a meaningful step toward a new paradigm in cross-modal retrieval with scalable generative multimodal embedding models.

\noindent\textbf{Acknowledgements.} 
This research is supported by EPSRC Programme Grant VisualAI EP$\slash$T028572$\slash$1, a Royal Society
Research Professorship RSRP$\backslash$R$\backslash$241003, 
and Scientific Research Innovation Capability Support Project for Young Faculty (ZYGXQNJSKYCXNLZCXM-I22).


\bibliographystyle{IEEEtran}
\bibliography{sample-base}

\end{document}